\documentclass[12pt]{article}%
\usepackage{standalone}
\usepackage{tikz}

\usepackage[nosort]{cite}
\usepackage{graphicx}
\usepackage{multicol}
\usepackage{amsfonts}
\usepackage{amssymb}
\usepackage{amsmath}
\usepackage{heck}
\usepackage{afterpage}
\usepackage{setspace}
\usepackage{verbatim}
\usepackage{color}
\usepackage{longtable}
\usepackage{subcaption}
\usepackage{epsfig}
\usepackage{epstopdf}
\usepackage{adjustbox}
\usepackage[margin=1in]{geometry}
\usepackage{titletoc}
\usepackage{hyperref}%
\usepackage{mathrsfs}
\usepackage{pgfplots}
\usepackage{caption}
\usepackage{float}
\usepackage{makecell}
\usepackage{mathtools}
\usepackage{pdflscape}
\usepackage{array}
\usepackage{braket}
\usepackage{multirow}
\usepackage{physics}
\newcolumntype{C}{>{\centering\arraybackslash}m{2cm}}

\numberwithin{equation}{section}
\newsavebox{\mysavebox}
\newlength{\myrest}
\hypersetup{
  colorlinks=true,
  linkcolor=blue,
  citecolor=blue,
  filecolor=blue,
  urlcolor=blue,
  bookmarksopen=true,
  bookmarksnumbered=true,
  pdftitle={Topological Band Theory for High-Dimensional Phase Spaces},
  pdfauthor={Vivek Chakrabhavi, Walt Chavarria, Jonathan J. Heckman, and Steven Rayan}
}

\usetikzlibrary{decorations.markings}
\usetikzlibrary{hobby}
\usetikzlibrary{arrows}
\usetikzlibrary{arrows.meta}
\usetikzlibrary{arrows,decorations.pathmorphing}
\usetikzlibrary{shapes.geometric,calc,arrows, positioning,shapes.misc,decorations.markings}
\tikzset{
  big arrow/.style={
    decoration={markings,mark=at position 1 with {\arrow[scale=2,#1]{>}}},
    postaction={decorate},
    shorten >=0.4pt},
  big arrow/.default=black}
\usetikzlibrary{arrows.meta,positioning,calc}

\pgfdeclarelayer{edgelayer}
\pgfdeclarelayer{nodelayer}
\pgfsetlayers{edgelayer,nodelayer,main}
\pgfplotsset{compat=1.16}
\tikzstyle{none}=[inner sep=0pt]

\tikzstyle{NodeCross}=[draw, shape=circle, cross out, inner sep=0pt, minimum size=6pt,line width=0.25mm]
\tikzstyle{Circle}=[draw, shape=circle, black, fill=black, inner sep=0pt, minimum size=6pt]
\tikzstyle{circle}=[draw, shape=circle, black, fill=black, inner sep=0pt, minimum size=16pt]
\tikzstyle{Star}=[draw, shape=star, fill=black, star points=8, inner sep=0pt, minimum size=8pt]
\tikzstyle{CircleRed}=[draw, shape=circle, black, fill=red, inner sep=0pt, minimum size=6pt]
\tikzstyle{StarP}=[draw={rgb,255: red,128; green,0; blue,128}, shape=star, fill={rgb,256: red,128; green,0; blue,128}, star points=8, inner sep=0pt, minimum size=12pt]
\tikzstyle{ShadedCircRed}=[draw=red, shape=circle, fill={rgb, 255: red,255; green,114; blue, 118}, inner sep=0pt, minimum size=80pt, line width=0.5mm, fill opacity=0.2]
\tikzstyle{ShadedCircRed2}=[draw=red, shape=circle, fill={rgb, 255: red,255; green,114; blue, 118}, inner sep=0pt, minimum size=10pt]
\tikzstyle{ShadedCircRed3}=[draw=black, shape=rectangle, fill={rgb, 255: red,255; green,114; blue, 118}, inner sep=0pt, minimum size=113pt, line width=0.25mm]
\tikzstyle{ShadedCirc}=[draw=red, shape=circle, fill=white, inner sep=0pt, minimum size=45pt,  fill opacity=1.0,  line width=0.5mm]
\tikzstyle{CircleBlue}=[draw, shape=circle, fill=blue, inner sep=0pt, minimum size=6pt]
\tikzstyle{BigCirclePurple}=[draw, shape=circle, fill={rgb,255: red,191; green,0; blue,191}, inner sep=0pt, minimum size=12pt]
\tikzstyle{CirclePurple}=[draw, shape=circle, fill={rgb,255: red,191; green,0; blue,191}, inner sep=0pt, minimum size=5pt]
\tikzstyle{EmptyCircle}=[draw, shape=circle, inner sep=0pt, minimum size=4pt]
\tikzstyle{GreenCircle}=[draw, shape=circle,  fill={rgb,255: red,80; green,200; blue,120}, inner sep=0pt, minimum size=8pt]
\tikzstyle{BrownCircle}=[draw, shape=circle,  fill={rgb,255: red,210; green,105; blue,30}, inner sep=0pt, minimum size=8pt]
\tikzstyle{CirclePurpleSmall}=[draw, shape=circle, fill={rgb,255: red,191; green,0; blue,191}, inner sep=0pt, minimum size=4pt]
\tikzstyle{BigCircleGreen}=[draw, shape=circle, fill={rgb,255: red,0; green,191; blue,0}, inner sep=0pt, minimum size=12pt]
\tikzstyle{BigCircleBlue}=[draw, shape=circle, fill={rgb,255: red,0; green,0; blue,191}, inner sep=0pt, minimum size=12pt]
\tikzstyle{BigCircleRed}=[draw, shape=circle, fill={rgb,255: red,191; green,0; blue,0}, inner sep=0pt, minimum size=12pt]
\tikzstyle{BrownCircleSmall}=[draw, shape=circle,  fill={rgb,255: red,210; green,105; blue,30}, inner sep=0pt, minimum size=6pt]
\tikzstyle{SmallCircleBrown}=[draw, shape=circle,  fill={rgb,255: red,210; green,105; blue,30}, inner sep=0pt, minimum size=5pt]
\tikzstyle{SmallCircleRed}=[draw, shape=circle, fill={rgb,255: red,191; green,0; blue,0}, inner sep=0pt, minimum size=6pt]
\tikzstyle{DashedLine}=[-, densely dashed, line width=0.25mm]
\tikzstyle{DottedLine}=[-, dotted, line width=0.25mm]
\tikzstyle{ThickLine}=[-, line width=0.25mm]
\tikzstyle{ArrowLineRight}=[-, -{Stealth[scale=1.25]}, line width=0.25mm, scale=5]
\tikzstyle{ArrowLineRed}=[-, draw={rgb,255: red,191; green,0; blue,0}, -{Stealth[scale=1.75]}, line width=0.1mm, scale=5]
\tikzstyle{RedLine}=[-, draw={rgb,255: red,191; green,0; blue,0}, fill=none, line width=0.5mm]
\tikzstyle{DashedLineThin}=[-, densely dashed, line width=0.125mm, fill=none, draw=black]
\tikzstyle{DottedRed}=[-, dotted, draw={rgb,255: red,191; green,0; blue,0}, fill=none, line width=0.25mm]
\tikzstyle{DashedRed}=[-, densely dashed, draw={rgb,255: red,191; green,0; blue,0}, fill=none, line width=0.25mm]
\tikzstyle{BlueLine}=[-, draw={rgb,255: red,0; green,0; blue,191}, fill=none, line width=0.5mm]
\tikzstyle{ArrowLineBlue}=[-, draw={rgb,255: red,0; green,0; blue,191}, -{Stealth[scale=1.75]}, line width=0.1mm, scale=5]
\tikzstyle{GreenDoubleArrow}=[<->, draw={rgb,155: red,0; green,255; blue,0},  line width= 0.5mm, scale=5]
\tikzstyle{RedDoubleArrow}=[<->, draw={rgb,255: red,255; green,0; blue,0},  line width= 0.5mm, scale=5]
\tikzstyle{BlueDottedLight}=[-, dotted, draw={rgb,255: red,0; green,0; blue,191}, fill=none, line width=0.3mm]
\tikzstyle{BrownLine}=[-, draw={rgb,255: red,210; green,105; blue,30}, fill=none, line width=0.5mm]
\tikzstyle{DottedRed}=[-, dotted, draw={rgb,255: red,191; green,0; blue,0}, fill=none, dotted, line width=0.5mm]
\tikzstyle{DottedPurple}=[-, dotted, draw={rgb,255: red,191; green,0; blue,191}, fill=none, dotted, line width=0.5mm]
\tikzstyle{BlueDottedLight}=[-, dotted, draw={rgb,255: red,0; green,0; blue,191}, fill=none, line width=0.5mm]
\tikzstyle{ArrowLinePurple}=[-, draw={rgb,255: red,191; green,0; blue,191}, -{Stealth[scale=1.75]}, line width=0.5mm, scale=5]
\tikzstyle{DashedLineGreen}=[-, densely dashed, draw={rgb,255: red,74; green,103; blue,65}, line width=0.25mm]
\tikzstyle{LineGreen}=[-, draw={rgb,255: red, 74; green,200; blue,65}, line width=0.5mm]
\tikzstyle{ArrowLineGreen}=[-, draw={rgb,255: red,0; green,191; blue,0}, -{Stealth[scale=1.75]}, line width=0.5mm, scale=5]
\tikzstyle{GreenLine}=[-, draw={rgb,255: red,0; green,191; blue,0}, fill=none, line width=0.5mm]
\tikzstyle{PurpleLine}=[-, draw={rgb,255: red,191; green,0; blue,191}, fill=none, line width=0.5mm]
\tikzstyle{PPurpleLine}=[-, draw={rgb,255: red,191; green,0; blue,191}, fill=none, line width=2.5mm]
\tikzstyle{DPurpleLine}=[-, dotted, draw={rgb,255: red,191; green,0; blue,191}, fill=none, line width=0.5mm]
\tikzstyle{SBrownLine}=[-, draw={rgb,255: red,191; green,0; blue,191}, fill=none, opacity=0.35, line width=2.5mm]
\tikzstyle{DottedBlue}=[-, dotted, draw=blue, fill=none, dotted, line width=0.5mm]
\tikzstyle{DashedPurpleLine}=[-, densely dashed, draw={rgb,255: red,191; green,0; blue,191}, fill=none, line width=0.5mm]
\tikzstyle{SmallCircleBlue}=[draw, shape=circle, fill=blue, inner sep=0pt, minimum size=5pt]
\tikzstyle{SmallCirclePurple}=[draw, shape=circle, fill={rgb,255: red,191; green,0; blue,191}, inner sep=0pt, minimum size=5pt]
\tikzset{snake it/.style={decorate, decoration=snake}}
\newcommand{\greatcirclearc}[6][]{%
  \pgfmathsetmacro{\gcxone}{#2}
  \pgfmathsetmacro{\gcyone}{#3}
  \pgfmathsetmacro{\gcxtwo}{#4}
  \pgfmathsetmacro{\gcytwo}{#5}
  \pgfmathsetmacro{\gcR}{#6}
  \pgfmathsetmacro{\gczone}{sqrt(max(\gcR*\gcR-\gcxone*\gcxone-\gcyone*\gcyone,0))}
  \pgfmathsetmacro{\gcztwo}{sqrt(max(\gcR*\gcR-\gcxtwo*\gcxtwo-\gcytwo*\gcytwo,0))}
  \pgfmathsetmacro{\gcdot}{(\gcxone*\gcxtwo+\gcyone*\gcytwo+\gczone*\gcztwo)/(\gcR*\gcR)}
  \pgfmathsetmacro{\gctheta}{acos(\gcdot)}
  \def\gcpoints{}
  \foreach \t in {0,0.02,...,1}{
    \pgfmathsetmacro{\gca}{sin((1-\t)*\gctheta)/sin(\gctheta)}
    \pgfmathsetmacro{\gcb}{sin(\t*\gctheta)/sin(\gctheta)}
    \pgfmathsetmacro{\gcxt}{\gca*\gcxone+\gcb*\gcxtwo}
    \pgfmathsetmacro{\gcyt}{\gca*\gcyone+\gcb*\gcytwo}
    \xdef\gcpoints{\gcpoints (\gcxt,\gcyt)}
  }
  \draw[#1] plot[smooth] coordinates {\gcpoints};
}

\tikzset{
  mid arrow/.style={
    postaction={
      decorate,
      decoration={
        markings,
        mark=at position 0.5 with {
          \arrow{Stealth[length=1.5mm,width=1.2mm]}
        }
      }
    }
  }
}
\tikzset{
dashstar/.style={
 dash pattern=on 5pt off 5pt,
 postaction={
  decorate,
  decoration={
   markings,
   mark=between positions 9pt and 1 step 10pt with {
     \node[color=red] {*};
   }
  }
 }
},
dashstarstar/.style={ 
 dash pattern=on 5pt off 10pt,
 postaction={
   decorate,
   decoration={
     markings,
     mark=between positions 10pt and 1
          step 15pt
           with {
            \node at (-2pt,0pt) {\pgfuseplotmark{star}};
            \node at (2pt,0pt) {\pgfuseplotmark{star}};
           }
   }
 }
}
}
\usetikzlibrary{decorations.markings, arrows.meta}
\tikzset{
    singlearrow/.style={
        postaction={decorate},
        decoration={markings, mark=at position 0.5 with {\arrow{Stealth}}}
    },
    doublearrow/.style={
      postaction={decorate},
      decoration={markings,
        mark=at position 0.42 with {\arrow{Stealth}},
        mark=at position 0.58 with {\arrow{Stealth}}}
    }
}

\newcommand{\g}[5]{\draw (#1,#2) arc[start angle=#3,end angle=#4,radius=#5];}

\begin{document}
\hypersetup{pageanchor=false}

\date{September 2026}

\title{Topological Band Theory for \\[4mm] High-Dimensional Parameter Spaces}

\let\originalPUTeXinstitutions\PUTeXinstitutions
\renewcommand{\PUTeXinstitutions}{{\footnotesize\originalPUTeXinstitutions}}
\institution{PENN}{\centerline{$^{1}$Department of Physics and Astronomy, University of Pennsylvania, Philadelphia, PA 19104, USA}}
\institution{PENNLPS}{\centerline{$^{2}$College of Liberal and Professional Studies, School of Arts and Sciences, University of Pennsylvania, Philadelphia, PA 19104, USA}}
\institution{PENNmath}{\centerline{$^{3}$Department of Mathematics, University of Pennsylvania, Philadelphia, PA 19104, USA}}
\institution{SASKA}{\centerline{$^4$Department of Mathematics and Statistics, University of Saskatchewan, Saskatoon, SK, Canada}}
\institution{SASKAQuanta}{\centerline{$^5$Centre for Quantum Topology and Its Applications (quanTA), University of Saskatchewan, Saskatoon, SK, Canada}}

\authors{
Vivek Chakrabhavi\worksat{\PENN}\footnote{e-mail: \href{mailto:vivekcm@sas.upenn.edu}{\texttt{vivekcm@sas.upenn.edu}}},
Walt Chavarria\worksat{\PENNLPS}\footnote{e-mail: \href{mailto:wchavar@sas.upenn.edu}{\texttt{wchavar@sas.upenn.edu}}}, \\[4mm]
Jonathan J. Heckman\worksat{\PENN,\PENNmath}\footnote{e-mail: \href{mailto:jheckman@sas.upenn.edu}{\texttt{jheckman@sas.upenn.edu}}}, and
Steven Rayan\worksat{\SASKA,\SASKAQuanta}\footnote{e-mail: \href{mailto:rayan@math.usask.ca}{\texttt{rayan@math.usask.ca}}}
}

\abstract{\noindent We study the topological band theory of quantum Hall systems in which the adiabatic flux parameters of an external gauge potential form a high-dimensional manifold. This situation arises both for strongly correlated matter on spatial Riemann surfaces of genus $g > 1$, as well as in the formal study of matter on compact oriented manifolds of dimension $4\ell+2 = 2p$ coupled to $p$-form gauge potentials. In these cases, the underlying topological invariants defined over the parameter space are significantly richer than the low-dimensional (i.e., two-dimensional) case and involve both the curvature (i.e., first Chern class) as well as higher order curvature invariants. These higher curvature invariants correspond to topologically protected contributions to Kubo-like formulae, constructed from correlation functions of the physical current operators. We illustrate these general considerations with an explicit example from hyperbolic band theory based on the genus-two Bolza Riemann surface, a case which has recently been simulated on a synthetic-dimension platform.}

\maketitle

\clearpage
\hypersetup{pageanchor=true}
\pagenumbering{roman}

\enlargethispage{\baselineskip}

\setcounter{tocdepth}{3}

\tableofcontents

\clearpage
\pagenumbering{arabic}

\section{Introduction}

Following the experimental discovery of the quantum Hall effect \cite{Early_Klitzing1980, Early_Laughlin1981}, much work was dedicated to a theoretical construction of the phenomenon, resulting in the celebrated TKNN formula for the quantum Hall conductivity \cite{Disorder_AokiAndo1981, Disorder_Prange1981, Edge_Halperin1982, TKNN_Thouless1982, TKNN_Avron1983, TKNN_Simon1983, TKNN_Berry1984, TKNN_Kohmoto1985, Torus_Avron1985, Torus_Niu1985, Edge_Hatsugai1993}, with Hall conductivity $\sigma_{xy} = \sigma_{\mathrm{Hall}} = \nu / 2 \pi$ controlled by the Chern class of a Berry connection defined on momentum space:
\begin{equation}\label{eq:BASIC}
\nu = \int_{\widetilde{T}^2} \frac{\mathcal{F}}{2\pi } = \int_{\widetilde{T}^2} c_1(\mathcal{L}) \in \mathbb{Z},
\end{equation}
where $\widetilde{T}^2$ is the two-torus of the accompanying Brillouin zone and $c_1(\mathcal{L}) \equiv \mathcal{F} / 2\pi $
is a Berry curvature defined over the Brillouin zone.

Importantly, the conductivity is a topological invariant, and thus topologically protected from small variations of parameters. These discoveries sparked the field of topological band theory \cite{TBT1, TBT2, TBT3, TBT4, TBT5, TBT6, TBT7, TBT8, TBT9, TBT10, TBT11, TBT12, TBT13, TBT14, TBT15, TBT16, TBT17, TBT18, TBT19, TBT20, TBT21, TBT22, TBT23, TBT24, TBT25, TBT26}. In many cases of interest, one works with a periodic tiling on a spatial $T^2$ in which the structure of the fundamental unit cell dictates non-trivial topological structure on the Brillouin zone.

Recent work has broadened topological band theory beyond Euclidean crystals and ordinary crystal momentum. This situation naturally occurs by considering the same underlying physical system on a genus $g > 1$ Riemann surface. Such systems can be synthetically realized on designer circuits \cite{Kollar2019Hyperbolic}. In this setting, hyperbolic band theory replaces the abelian translation lattice by a cocompact Fuchsian group. Its irreducible representations need not be one-dimensional, so in mathematical terms the analogue of a Brillouin zone naturally includes moduli of non-abelian representations or, geometrically, vector bundles on a higher-genus surface \cite{Rayan2021Hyperbolic, Rayan2022Hyperbolic, Rayan2022Higgs}.\footnote{In more detail, the hyperbolic Bloch transform gives an analytic realization of this representation-theoretic decomposition and, in its geometric form, sends wavefunctions to sections of stable flat bundles on the compact quotient \cite{NagyRayan2024}. In the untwisted rank-one sector, the character torus $\operatorname{Hom}(\pi_1(\Sigma_g),U(1))\cong H^1(\Sigma_g;\mathbb R)/H^1(\Sigma_g;\mathbb Z)$ is precisely the abelian flux torus used below. Recent calculations of hyperbolic Chern-insulator response make this connection especially direct \cite{TBT26}. At the same time, stable $K$-classes can forget geometric and holomorphic variation over these sector spaces even when that variation still controls spectra, Berry holonomy, the quantum metric, or partially filled response \cite{Rayan2026BeyondKTheory}. This observation motivates retaining the differential-geometric data on the parameter space rather than only its stable $K$-theory class. Related departures from an ordinary Brillouin torus occur in quasiperiodic systems, where phason parameters can supply extra dimensions and Chern numbers \cite{Kraus2012Quasicrystals}, in synthetic-dimension platforms, where internal or frequency modes emulate additional lattice directions \cite{Ozawa2017Synthetic, Lustig2019Synthetic}, and in non-Hermitian systems, where the skin effect can require a generalized complex Brillouin zone \cite{Yao2018NonHermitian}.} These developments share a common lesson: the topology relevant to response is often carried by a parameter or moduli space that is not simply the momentum torus of a Euclidean crystal.

With this in mind, our aim in this work will be to provide a natural generalization of equation (\ref{eq:BASIC}) to situations
where the parameter space for the gauge potential sweeps out a space of dimension $d > 2$.
This occurs in the case of gapped systems on a genus $g > 1$ Riemann surface, and also arises in higher-dimensional generalizations
of quantum Hall systems in which the degrees of freedom couple to a $p$-form potential on a manifold $X_{4 \ell + 2}$ of dimension $4\ell + 2 = 2p$ (see e.g., \cite{Heckman:2017uxe}).\footnote{For different approaches to integer and fractional quantum Hall systems in higher spacetime dimensions, see e.g., \cite{ZhangHu:2001, KarabaliNair:2002, Heckman:2017uxe, Tong:2018}.}

The quantum Hall system couples to such a $p$-form potential via a perturbation of the Hamiltonian:
\begin{equation}
\Delta H = - \int_{X} \mathrm{dVol}_{X} \, \frac{1}{p!} J_{i_1 ... i_p} A^{i_1 ... i_p} =  \int_{X} \ast_{X} J_p \wedge A_p,
\end{equation}
where $\mathrm{dVol}_{X}$ is the volume form on $X$, $J_{i_1 ... i_p}$ is a fully anti-symmetric current and $A^{i_1 ... i_p}$
implicitly defines a $p$-form gauge potential which we write as the differential form $A_p$.

Even in the case $p = 1$, generalizing to a $g > 1$ Riemann surface leads to a high-dimensional parameter space. Indeed, the moduli
are captured by the holonomies, i.e., the gauge invariant periods of $A_1$ obtained by integrating over a basis of 1-cycles $C_M$ of $X$:
\begin{equation}
2 \pi \theta^{M} \equiv \int_{C_M} A_1.
\end{equation}
Observe that changing their values amounts to varying the ground state wavefunction and so we can construct a Berry connection on this parameter space, $\mathcal{M}_{\Theta}$. In the special case of a genus $g = 1$ Riemann surface, $\mathcal{M}_{\Theta}$ is again just a two-torus, and it is topologically the same torus as that provided by the Brillouin zone $\widetilde{T}^{2}$. More generally, however, $\mathcal{M}_{\Theta}$ is a higher-dimensional geometry, i.e., it is the Jacobian of $X$,\footnote{As follows from the Abel-Jacobi map. See reference \cite{acgh1985geometry} for a review.} which is $T^{2g}$. This is to be expected: we have $2g$ different 1-cycles on $X$ and each period gives a real parameter.

Similar considerations apply in higher-dimensional generalizations of the quantum Hall effect for $X_{4 \ell + 2}$ with $\ell > 0$. Indeed, the only change is that now we integrate the $p$-form potential $A_{p}$ over middle-dimensional cycles $C_M$ of $X_{4 \ell + 2}$:
\begin{equation}
2 \pi \theta^{M} \equiv \int_{C_M} A_p.
\end{equation}
Varying the ground state wavefunction with respect to these parameters again allows us to construct a Berry connection on the parameter space $\mathcal{M}_{\Theta}$. A priori, this could be a high-dimensional space, and is generally not two-dimensional.

Much as in the standard setting, we can still construct a Berry curvature $\mathcal{F}_{MN}$,
i.e., a formal two-form on $\mathcal{M}_{\Theta}$. Observe, however,
that when $\mathrm{dim} \mathcal{M}_{\Theta} > 2$, we cannot simply integrate $\mathcal{F}$ over $\mathcal{M}_{\Theta}$. This
motivates a basic question: what is the natural generalization of equation (\ref{eq:BASIC}) to this 
broader setting?

Our basic claim is that the Hall conductivity in this case is:
\begin{equation}
\sigma_{\mathrm{Hall}} = \frac{1}{2 \pi} \int_{\mathcal{M}_{\Theta}} S^{d-1} \wedge \frac{\mathcal{F}}{2\pi} = \frac{1}{2 \pi}\int_{\mathcal{M}_{\Theta}} [S]^{d-1} \smile c_1(\mathcal{L}) \in \frac{1}{2\pi}\mathbb{Z},
\end{equation}
where the dimension of the moduli space is related to $d$ via $2d = \mathrm{dim} \mathcal{M}_{\Theta}$,
and $S$ is a quantized symplectic form on $\mathcal{M}_{\Theta}$ implicitly defined by the intersection pairing on
$X_{4\ell+2}$. The righthand side is manifestly an integer invariant multiplied by $1 / 2 \pi$, and as such directly
produces a quantized TKNN invariant.

This is but the first of a sequence of topological invariants available to us in the higher-dimensional setting:
\begin{equation}
\nu_{m} = \int_{\mathcal{M}_{\Theta}} S^{d-m} \wedge \left(\frac{\mathcal{F}}{2\pi } \right)^m \in \mathbb{Z},
\end{equation}
for $m = 1,...,d$. These additional integral invariants naturally appear in higher-order (disconnected) current correlation functions.
In general, they are distinct numbers and do not directly follow from the data of $\sigma_{\mathrm{Hall}}$ alone.

The rest of this paper is organized as follows. We present a brief review of topological band theory for quantum Hall systems on a
Euclidean $T^2$ in section \ref{sec:WARMUP}. In section \ref{sec:BERRY} we construct the Berry connection obtained from varying an external gauge field. With this in place, we show that this leads to a class of generalized Hall conductivities in section \ref{sec:QUANTIZATION}.
Section \ref{sec:BOLZA} presents an explicit example based on the genus-two Bolza surface. We present our conclusions and potential directions for future investigation in section \ref{sec:CONC}. Some additional details on quantized symplectic forms are given in Appendix \ref{app:LEFSCHETZ}. We discuss some further candidate invariants based on the Chern-Simons form constructed from the Berry connection in Appendix \ref{app:CS}.
Additional details on the geometry of the Bolza surface are given in Appendix \ref{app:BOLZA}.

\section{Review of Euclidean \texorpdfstring{$T^2$}{T2} Case} \label{sec:WARMUP}

In this section we briefly review the special case where our gapped system is placed on a spatial
$T^2$. With this in place, we generalize to higher genus and higher-dimensional settings in section \ref{sec:BERRY}.

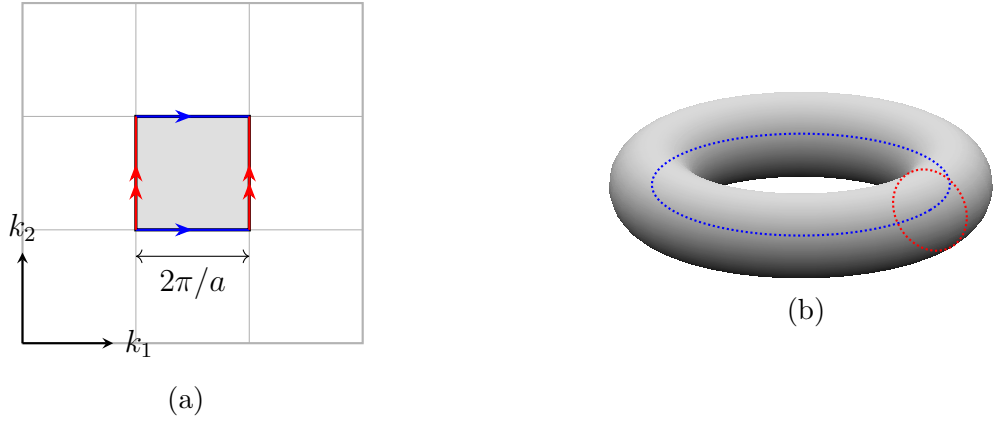
\begin{figure}[t!]
\centering

\begin{subfigure}[c]{0.45\textwidth}
\centering
\begin{tikzpicture}[scale=1]
  \def\s{1.5} 
  
  \draw[gray!60, step=\s] (0,0) grid (3*\s,3*\s);
  \draw[gray!60, thick] (0,0) rectangle (3*\s,3*\s);

  \draw[very thick, fill=gray!25] (\s,\s) rectangle (2*\s,2*\s);

  \draw[singlearrow, thick, blue] (\s,2*\s) -- (2*\s,2*\s);
  \draw[singlearrow, thick, blue] (\s,\s)   -- (2*\s,\s);

  \draw[doublearrow, thick, red] (\s,\s)   -- (\s,2*\s);
  \draw[doublearrow, thick, red] (2*\s,\s) -- (2*\s,2*\s);

  \draw[<->] (\s,\s-0.35) -- (2*\s,\s-0.35)
      node[midway, below] {$2\pi/a$};

\draw[->, thick, >=stealth] (0,0) -- (0.8*\s, 0) node[right] {$k_1$};
\draw[->, thick, >=stealth] (0,0) -- (0, 0.8*\s) node[above] {$k_2$};
\end{tikzpicture}
\label{fig:torus-square}
\caption{}
\end{subfigure}
\hspace{0.5cm}
\begin{subfigure}[c]{0.45\textwidth}
\centering
\begin{tikzpicture}
\begin{axis}[
  hide axis,
  view={30}{20},
  axis equal image,
  width=8.5cm,
  height=8.5cm,
  colormap={grayscale}{gray(0cm)=(0.0); gray(1cm)=(0.85)},
  z buffer=sort,
]

\addplot3[
  surf,
  shader=interp,
  domain=0:360,
  samples=40,
  y domain=0:360,
  samples y=40,
  opacity=0.95,
] ({(3.5+1*cos(y))*cos(x)}, {(3.5+1*cos(y))*sin(x)}, {1*sin(y)});

\draw[densely dotted, thick, blue] plot[variable=\t, domain=0:360, samples=120]
  (axis cs:{3.5*cos(\t)},{3.5*sin(\t)},{0});

\draw[densely dotted, thick, red] plot[variable=\t, domain=0:360, samples=120]
  (axis cs:{3.5+1*cos(\t)},{0},{1*sin(\t)});

\end{axis}
\end{tikzpicture}
\label{fig:torus-3d}
\caption{}
\end{subfigure}

\caption{Two representations of the Brillouin zone of a square lattice with spacing $a$. (a) A fundamental momentum-space cell of side $2\pi/a$, with opposite sides identified. (b) A schematic embedding of the resulting quotient torus. The radii of the embedded torus have no physical significance.}
\label{fig:torus}
\end{figure}

To begin, consider a gapped noninteracting system on a two-dimensional periodic lattice.  Translation invariance makes crystal momentum periodic, so the Brillouin zone is a two-torus.  Explicitly, for a square lattice with spacing $a$, each momentum component obeys
\begin{align}
    k_i \sim k_i + \frac{2\pi n}{a}, \hspace{0.1 cm} \text{for} \hspace{0.1 cm} n\in \mathbb{Z}. \label{momentum identification}
\end{align}
If we quotient $\mathbb{R}^2$ by the lattice formed under this identification, it follows that the momentum $(k_1,k_2)$ naturally lives on $T^2$ (see \hyperref[fig:torus]{Figure 1}).

Treating momentum as an adiabatic parameter, an isolated non-degenerate band defines a complex line bundle over the Brillouin torus.  If $\ket{\psi_0(k)}$ is a smooth local choice of normalized state, parallel transport around a closed path $C$ produces the Berry phase \cite{TKNN_Berry1984}
\begin{align}
    \phi = i\oint_C \mathcal{A}_i dk^i,
\end{align}
and $\mathcal{A}_i$ is
\begin{align}
    \mathcal{A}_i = -i\braket{\psi_0}{\frac{\partial \psi_0}{\partial k^i}}.
\end{align}
Thus $\mathcal{A}=\mathcal{A}_i\,dk^i$ is a local connection one-form on the band line bundle.  In a many-body treatment, the analogous line bundle is formed by a unique gapped ground state over a torus of boundary twists or inserted fluxes.

Next, defining the Hall conductivity via $\langle J_i \rangle = \sigma_{ij}E^j$, a linear response approach is taken to obtain the Kubo formula \cite{Kubo_Kubo1957, Kubo_Kubo1959, Kubo_Streda1982} for the Hall conductivity:
\begin{align}
    \sigma_{ij} = i\sum_{n \neq 0} \frac{\langle \psi_0 | J_j|\psi_n \rangle \langle \psi_n | J_i|\psi_0 \rangle - \langle \psi_0 | J_i|\psi_n \rangle \langle \psi_n | J_j|\psi_0 \rangle}{(E_n - E_0)^2},
\end{align}
where the $\ket{\psi_n}$ form a complete energy eigenbasis. The antisymmetric response is the Berry curvature, $\sigma_{ij}=\mathcal F_{ij}$, where
\begin{align}
    \mathcal{F}_{ij} = \frac{\partial \mathcal{A}_j}{\partial k^i} - \frac{\partial \mathcal{A}_i}{\partial k^j} = -i \left[ \braket{\frac{\partial\psi_0}{\partial k^i}}{\frac{\partial \psi_0}{\partial k^j}}  - \braket{\frac{\partial\psi_0}{\partial k^j}}{\frac{\partial \psi_0}{\partial k^i}}    \right].
\end{align}

With the convention used here, $[\mathcal{F}/(2\pi)]$ represents the first Chern class.  Hence its integral over the Brillouin torus is an integer $\nu \in \mathbb{Z}$. Restoring charge, Planck's constant, and any volume normalization converts this dimensionless Chern number into the physical Hall conductance:
\begin{equation}
\sigma_{\mathrm{Hall}} = \sigma_{xy} = \frac{\nu}{2 \pi}.
\end{equation}

\begin{figure}[t!]
\centering
\begin{tikzpicture}[baseline=(current bounding box.center)]
\begin{axis}[
  hide axis,
  view={30}{20},
  axis equal image,
  width=10cm,
  height=10cm,
  colormap={grayscale}{gray(0cm)=(0); gray(1cm)=(0.85)},
  z buffer=sort,
]

\addplot3[
  surf,
  shader=interp,
  domain=0:360,
  samples=40,
  y domain=0:360,
  samples y=40,
  opacity=0.95,
] ({(3.5+1*cos(y))*cos(x)}, {(3.5+1*cos(y))*sin(x)}, {1*sin(y)});

\draw[blue, solid, thick, postaction={
    decorate,
    decoration={
      markings,
      mark=at position 0.20 with {\arrow[thick]{Stealth}}, mark=at position 0.70 with {\arrow[thick]{Stealth}}
    }
  }] plot[variable=\t, domain=0:360, samples=120]
  (axis cs:{3.5*cos(\t)},{3.5*sin(\t)},{0});

\node[anchor=west] at (axis cs:4.3, 0, -0.5) {$A_1$};

\draw[red, solid, thick, postaction={
    decorate,
    decoration={
      markings,
      mark=at position 0.05 with {\arrow[thick]{Stealth}}, mark=at position 0.55 with {\arrow[thick]{Stealth}}
    }
  }] plot[variable=\t, domain=0:360, samples=120]
  (axis cs:{3.5+1*cos(\t)},{0},{1*sin(\t)});

\node[anchor=south west] at (axis cs:-3.5, 0, 0.2) {$A_2$};

\end{axis}
\end{tikzpicture}
\caption{A spatial $2$-torus is shown with a background gauge potential $A$ threaded through its two fundamental cycles.}
\label{fig:torus-with-fluxes}
\end{figure}
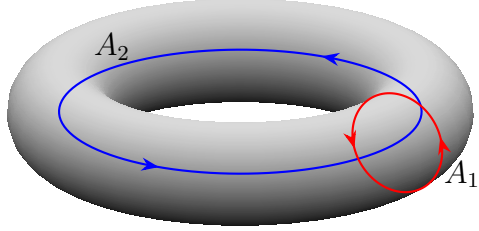

Rather than work directly in terms of the momentum space, we can instead opt to work with the parameter space defined by a background gauge field. On a spatial two-torus we can vary the two holonomies of a flat background $U(1)$ connection (see \hyperref[fig:torus-with-fluxes]{Figure 2}). Large gauge transformations identify each normalized holonomy modulo one, so the flux parameter space is itself a two-torus. Insofar as all matter couples via the covariant derivative:
\begin{equation}
\nabla_{j} = \partial_j + i A_j,
\end{equation}
we can opt to use either the momentum or the holonomies of the gauge field interchangeably and
the latter naturally generalizes to other situations. A unique gapped ground state over this flux torus again defines a Berry line bundle.  This construction does not identify the spatial torus $T^2$ with the parameter torus $\widetilde{T}^2$;
the two spaces arise for different reasons even though they happen to have the same dimension and topology in this example.

The geometry of the parameter space is the same as that of the Brillouin zone,
where now the moduli space of fluxes is the base space of a complex line bundle
associated with a principal $U(1)$-bundle endowed with the Berry connection.

The same adiabatic-response calculation identifies the antisymmetric response with the Berry curvature. Since this parameter space is two-dimensional, integrating the response two-form over the flux torus gives an integer after division by $2\pi$:
\begin{equation}
\nu = \int_{\widetilde{T}^2} \frac{\mathcal{F}}{2 \pi},
\end{equation}
with Hall conductivity given by:
\begin{equation}
\sigma_{xy} = \sigma_{\mathrm{Hall}} = \frac{\nu}{2 \pi}.
\end{equation}

In what follows we use this flux-space formulation because it accommodates interacting Hamiltonians and extends naturally to middle-degree gauge fields on higher-dimensional spatial manifolds.

The equality of dimensions in the ordinary two-dimensional example is a special feature.  For a genus-$g$ surface $\Sigma_g$, there are $2g$ independent continuous holonomies, so the connected flux torus has dimension $2g$.  For $g>1$, the Berry curvature is no longer a top form, and the preceding full-parameter-space integral is unavailable without further structure. The same is true for higher-dimensional generalizations of the quantum Hall effect. For example, consider the spatial manifold $T^6$ and thread a $3$-form gauge potential through the $3$-cycles. By counting the number of $3$-cycles, we conclude that the moduli space of fluxes has dimension $\binom{6}{3}= 20$.

Our task will therefore be to understand the geometry of the parameter space in its own terms, and explain how Chern characters constructed over this moduli space define physical observables associated with generalized Hall conductivities.

\section{Berry Connection and Generalized Conductivities} \label{sec:BERRY}

We now turn to gapped systems coupled to an external background $p$-form gauge potential
in which the parameter space for the gauge field is not a two-dimensional $\widetilde{T}^2$.
We work on a compact oriented spatial $4\ell+2$-dimensional Riemannian manifold $X_{4 \ell + 2}$. The spacetime $Y = \mathbb{R}_{\mathrm{time}} \times X_{4 \ell + 2}$ has dimension $4 \ell + 3$. We assume that in the absence of the external $p$-form field, we have a system of strongly correlated degrees of freedom with a mass gap, as governed by a many-body Hamiltonian $H$. We denote by $\vert \psi_0 \rangle$ the ground state for this unperturbed system. A priori, this background Hamiltonian could have many couplings and realize a rather intricate band structure. That being said, we shall mainly treat it as a strongly correlated ``black box''. Indeed, our primary interest will instead be in understanding the response of this system to a background field, as governed by switching on a perturbation to the Hamiltonian density of the form:
\begin{equation}
\Delta H = - \underset{X}{\int} \mathrm{d Vol}_X \, \frac{1}{p!}J_{i_1 ... i_p} \wedge A^{i_1 ... i_p} = \underset{X}{\int}  \ast_{X} J_p \wedge A_p,
\end{equation}
namely, we assume the degrees of freedom couple to a $p$-form potential $A_{p}$, and $J_{p}$ is the current associated with the degrees of freedom of the background. Here, $p = 2 \ell + 1$, meaning $\ast_{X} J_p$ is also a $p$-form. Observe that under a gauge transformation $A_{p} \mapsto A_{p} + d \lambda_{p-1}$, gauge invariance amounts to the condition that $d \ast_{X} J_{p} = 0$, i.e., current conservation.

The Hall conductivity is implicitly specified by the relation:
\begin{equation}
\langle J_{j_1 ... j_{p}} \rangle = \frac{1}{p!}\sigma_{j_1 ... j_{p} j_{p+1} ... j_{2p}} F^{0 j_{p+1} ... j_{2p}},
\end{equation}
where $F^{0 j_{p+1} ... j_{2p}}$ is the ``electric'' component of the $(p+1)$-form field strength $F_{p+1}$.
This generalized conductivity follows from a Kubo-like formula, much as in the ``standard case,'' namely:
\begin{align}
\sigma_{j_1 ... j_{2p}}  = i \sum_{r\ne0}
    \frac{\langle\psi_0| J_{[j_{p+1} ... j_{2p}]}|\psi_r\rangle
    \langle\psi_r| J_{[j_1 ... j_{p}]}|\psi_0\rangle - ([j_1...j_p] \leftrightarrow [j_{p+1} ... j_{2p}])}{(E_r-E_0)^2},
    \label{Hall Time Integral}
\end{align}
where $J_{[j_1...j_p]}$ are appropriate integrated current zero modes. Our interest is in the fully anti-symmetrized combination, which we refer to as the generalized Hall conductivity:
\begin{equation}
\sigma_{\mathrm{Hall}} = \frac{1}{(2 p)!} \varepsilon^{j_1 ... j_{2p}} \sigma_{j_1 ... j_{2p}}.
\end{equation}
This generalized Hall conductivity sets the level of a general Chern-Simons-like
theory on $Y_{4 \ell + 3} = \mathbb{R}_{\mathrm{time}} \times X_{4\ell+2}$
with action \cite{Heckman:2017uxe}:
\begin{equation}
\mathcal{S}_{\mathrm{gen-CS}} \equiv \frac{\nu}{4 \pi} {\int} A_{p} \wedge d A_{p} - \ast_{Y} J_{p} \wedge A_{p},
\end{equation}
where the last term corresponds to the coupling of the background field to the current.
The level $\nu$ is related to the Hall conductivity via (see e.g., \cite{Tong:2016kpv, Heckman:2017uxe}):
\begin{equation}
\sigma_{\mathrm{Hall}} = \frac{\nu}{2 \pi}.
\end{equation}
In particular, gauge invariance requires $\nu \in \mathbb{Z}$, i.e., quantization of the Hall conductivity.

Our aim will be to understand this quantization condition directly in terms of the topological band theory of the system.
To this end, we track the behavior of the ground state wavefunction as we vary the background gauge field $A_{p}$.
We keep the background flux $F_{p+1}$ (locally specified as $F_{p+1} = dA_{p}$) fixed, and instead consider the adiabatic variation from the holonomies, i.e., periods of $A_{p}$.

Along these lines, introduce a basis of middle dimensional cycles $C_{M} \in H_{2 \ell +1}(X ; \mathbb{Z})_{\mathrm{free}}$, i.e., we drop all torsion (namely discrete $\mathbb{Z} / k \mathbb{Z}$ type contributions). We track the response with respect to the holonomies:
\begin{equation}
2 \pi \theta^{M} \equiv \underset{C_{M}}\int A_{p}.
\end{equation}
Without loss of generality, we can normalize $A_{p}$ so that the $\theta^{M} \sim \theta^{M} +  1$ form a unit torus in
$b = 2d = h_{p}(X,\mathbb{Z})_{\mathrm{free}}$ dimensions.

We now study the response of the ground state wavefunction to a change in the background values of the $\theta^{M}$, so we will seek to study the family of Hamiltonians $H(\theta)$ and accompanying ground states $\vert \psi_{0}(\theta) \rangle$ obtained from such variations.

Choose closed $(2 \ell +1)$-forms $\gamma_M$ representing an integral basis of $H^{p}(X ; \mathbb{Z})_{\mathrm{free}}$
and middle-dimensional cycles $C_M$ satisfying $\int_{C_M}\gamma_N=\delta^M{}_N$.
We consider the connected component of flat compact $(2\ell+1)$-form backgrounds.  Its moduli space is
\begin{align}
    \mathcal M_{\Theta} = H^{2\ell+1}(X;\mathbb R)/H^{2\ell+1}(X;\mathbb Z)_{\mathrm{free}}
    \cong (\mathbb R/\mathbb Z)^b. \label{flux moduli}
\end{align}
With respect to this basis of differential forms, we can locally present our $p$-form gauge potential as:
\begin{equation}
A_p = 2 \pi \theta^{M} \gamma_{M},
\end{equation}
in the obvious notation. Likewise, we decompose the Hodge dual of the current $\ast_{X} J_{p}$ as:\footnote{Observe that the normalization
for $\theta^M$ and $\zeta^M$ is slightly different. The present choice leads to more natural integrality conditions later.}
\begin{equation}\label{eq:Jbasis}
\ast_{X} J_{p} = \zeta^{M} \gamma_{M}.
\end{equation}

With this in place, we now study the ground state wavefunction $\vert \psi_{0} (\theta) \rangle$
and how it changes under a variation of the $\theta^{M}$. Geometrically, the
ground states form a Hermitian line bundle $\mathcal L\to\mathcal M$.
In a local gauge its Berry connection and curvature are:\footnote{We work in conventions where the Berry connection is Hermitian.}
\begin{align}
    \mathcal A_M&= - i\braket{\psi_0}{\partial_M\psi_0}, \label{Berry Connection}\\
    \mathcal F&=d\mathcal A=\frac12\mathcal F_{M N}\,d\theta^{M}\wedge d\theta^{N},\\
    \mathcal F_{M N}&=\partial_M\mathcal A_N-\partial_N\mathcal A_M,
\end{align}
i.e.:
\begin{equation}
\mathcal{F}_{MN} = -i(\braket{\partial_M\psi_0}{\partial_N\psi_0}
    -\braket{\partial_N\psi_0}{\partial_M\psi_0}). \label{Berry Curvature}
\end{equation}
The local one-form $\mathcal A$ is gauge dependent, whereas $\mathcal F$ is global and:
\begin{equation}
\left[\frac{\mathcal{F}}{2\pi} \right] = c_1(\mathcal L).
\end{equation}

There is a natural analogue of the generalized Kubo formula equation (\ref{Hall Time Integral}), but now specified as a
two-form on $\mathcal{M}_{\Theta}$:\footnote{Observe that the Hodge star on a middle-dimensional form on $X$ is rather innocuous.}\textsuperscript{,}\footnote{Note that $\zeta_M$ is defined as the integrated current in the direction dual to $\gamma_M$, given by
$\zeta_M = \frac{1}{2\pi} \partial_M (\Delta H) = \int_X \ast_X J \wedge \gamma_M$

\label{fn:Integrated Current}}
\begin{align}
    \sigma_{M N}=i \sum_{r\ne0}
    \frac{\langle\psi_0|\mathcal \zeta_N|\psi_r\rangle
          \langle\psi_r|\mathcal \zeta_M|\psi_0\rangle - (M \leftrightarrow N)}{(E_r-E_0)^2}. \label{Hall Time IntegralMTHETA}
\end{align}
Compared with $\sigma_{\mathrm{Hall}}$, this object is defined over the parameter space.
That being said, we can relate the two expressions via:
\begin{equation}
\sigma_{\mathrm{Hall}} = \frac{1}{2}\sigma_{M N} \int_{X}  \gamma^{M} \wedge \gamma^{N}.
\end{equation}
The integration of $\gamma^{M} \wedge \gamma^{N}$ will be interpreted in section \ref{sec:QUANTIZATION}
in terms of an integral symplectic form defined on $\mathcal{M}_{\Theta}$.

The two-index tensor $\sigma_{M N}$ is closely related to the Berry curvature of our moduli space. To see why,
consider the change of the ground state as a function of the parameters $\theta^{M}$. First order perturbation theory reveals:
\begin{align}
    \ket{\delta_M\psi_0}
    =\sum_{r\ne0}\frac{\ket{\psi_r}\braket{\psi_r}{\partial_M\psi_0}}{E_r-E_0}.
\end{align}
On the other hand, our expression for the Berry curvature from line (\ref{Berry Curvature}) is:
\begin{equation}
\mathcal{F}_{M N} = -i \left(\braket{\partial_M\psi_0}{\partial_N \psi_0}
    -\braket{\partial_N \psi_0}{\partial_M \psi_0}\right).
\end{equation}
So, the Berry curvature is directly related to our generalized two-form on $\mathcal{M}_{\Theta}$:
\begin{equation}
2 \pi \sigma_{M N} = \frac{\mathcal{F}_{M N}}{2 \pi}.
\end{equation}

Compared with the case of a physical system defined on a spatial $T^2$, observe that $\mathcal{F}_{M N}$ is generally not a top-form on $\mathcal{M}_{\Theta}$. Our aim in the next section will be to show how $\mathcal{F}_{MN} / 2 \pi$ returns observable integers, generalizing the standard Hall conductivity to a collection of integral invariants, as specified by current correlation functions specified on the physical spacetime.

\section{Quantization of Generalized Hall Conductivities} \label{sec:QUANTIZATION}

In the previous section we arrived at a generalized Kubo-like formula for a conductivity $\sigma_{MN}$ defined
as a two-form on $\mathcal{M}_{\Theta}$. Our aim in this section will be to show on general grounds that this leads (upon suitable integration) to a quantized Hall conductivity, and moreover, that it leads to a collection of integral invariants which naturally generalize the
special case of the standard TKNN invariant.

As already mentioned, the issue is that only for a two-dimensional flux torus is the Berry curvature a top form, allowing the conclusion:
\begin{align}
    \int_{\mathcal{M}_{\Theta}}\frac{\mathcal F}{2\pi}=C\in\mathbb Z. \label{First Chern number}
\end{align}
When $\dim\mathcal{M}_{\Theta}>2$, one may still integrate $\mathcal F/ 2\pi$ over any integral two-cycle.
In particular, the coordinate subtori give integers
\begin{align}
    C_{MN}=\int_{K_{MN}}\frac{\mathcal F}{2\pi}\in\mathbb Z, \label{2 dim. Chern numbers}
\end{align}
where $K_{MN}$ is the coordinate subtorus obtained by varying $\theta^M$ and $\theta^N$. To obtain a distinguished number involving the entire flux torus, we now use the middle-dimensional intersection pairing of $X$ as additional geometric input. The construction applies
whenever $\mathcal{M}_{\Theta}$ is specified by a higher-dimensional torus. Some helpful examples to keep in mind are $X$ given by genus-$g$ Riemann surfaces and higher-dimensional tori, though we emphasize that the construction applies far more broadly.

\subsection{The Middle-Dimensional Intersection Lattice}

On the free lattice $\Lambda_X = H^{p}(X;\mathbb{Z})_{\mathrm{free}}$,
Poincar\'e duality defines the alternating unimodular form
\begin{align}
    Q_{MN}=\int_X\gamma_M\wedge\gamma_N. \label{intersection integral}
\end{align}
An alternating unimodular lattice has even rank $b=2d$ and admits an integral symplectic
basis \cite{Hatcher2002, Bredon1993}. We order it as
\begin{align}
    (\gamma_1,\ldots,\gamma_{2d})
    =(\alpha_1,\ldots,\alpha_d,\beta_1,\ldots,\beta_d),
\end{align}
with
\begin{equation}\label{symplectic basis}
\int_X\alpha_m\wedge\alpha_n=0,\qquad
\int_X\beta_m\wedge\beta_n=0,\qquad
\int_X\alpha_m\wedge\beta_n=\delta_{mn}.
\end{equation}
In this ordering,
\begin{align}
    Q=\begin{pmatrix}0&I_d\\-I_d&0\end{pmatrix}. \label{canonical Q}
\end{align}

\subsection{Construction of the Moduli Symplectic Form} \label{sec:3.2}

Because the normalized coordinates $\theta^M$ have unit period, $[d\theta^M]$ form an integral basis of $H^1(\mathcal M_{\Theta};\mathbb Z)$. Equation \eqref{intersection integral} therefore induces the constant two-form
\begin{align}
    S=\frac12Q_{MN}\,d\theta^M\wedge d\theta^N. \label{moduli symplectic form}
\end{align}
It is closed, non-degenerate, and integral. Thus the connected abelian flux moduli space is the smooth symplectic torus $(\mathcal{M}_{\Theta},S)$.

It is useful to separate this canonical construction from a more general prequantum choice.
A prequantum structure consists of a symplectic form $\omega$ together with a Hermitian
line bundle $\mathcal K$ and connection satisfying \cite{Kostant1970Quantization}:
\begin{align}
    F_{\nabla^{\mathcal K}}=-2\pi i\,\omega,
    \qquad [\omega]=c_1(\mathcal K)\in H^2(\mathcal{M}_{\Theta};\mathbb Z). \label{prequantum curvature}
\end{align}
The line bundle by itself fixes only the cohomology class, not a particular symplectic representative. Relative to the coordinates above, the class of any constant integral symplectic form is represented by a non-degenerate integral skew matrix $B$:
\begin{align}
    [\omega]=\frac12B_{MN}[d\theta^M]\smile[d\theta^N].
\end{align}
Since $Q$ is unimodular, $R=Q^{-1}B$ is an integral endomorphism of $\Lambda_X$, and
\begin{align}
    B_{MN}=Q_{MK}R^K{}_N
    =\int_X \gamma_M\wedge R(\gamma_N). \label{prequantum intersection relation}
\end{align}
Skew symmetry of $B$ is equivalent to $R^{t}Q=QR$,\footnote{Here, $R^{t}$ denotes the transpose.}
and non-degeneracy of $B$ is equivalent to invertibility of $R$ over $\mathbb R$.

A linear torus automorphism determined by $g\in\operatorname{GL}(2d,\mathbb Z)$ sends $Q$ to $g^{t}Qg$.
Every non-degenerate integral skew matrix has a normal form
\begin{align}
    U^tBU=\begin{pmatrix}0&\Delta\\-\Delta&0\end{pmatrix},
    \qquad \Delta=\operatorname{diag}(\Delta_1,\ldots,\Delta_d),
    \qquad \Delta_1\mid\cdots\mid \Delta_d,
\end{align}
whose elementary divisors $(\Delta_1,\ldots,\Delta_d)$ are its polarization type \cite{Latschev2013Gromov}.\footnote{Here $\Delta_1\mid\cdots\mid \Delta_d$ denotes the divisibility chain $\Delta_i\mid \Delta_{i+1}$.} Consequently, $B=g^tQg$ for an integral torus automorphism if and only if $B$ is principally polarized, that is, all $\Delta_i=1$.

This also clarifies the limits of a local Darboux argument. At any point, two symplectic forms can be put into the same standard matrix by a general change of tangent-space basis. A matrix that is symplectic with respect to $Q$, however, obeys $g^tQg=Q$ and cannot produce a different form. Pointwise Darboux equivalence neither identifies the integral period lattices nor glues the local changes of coordinates into a global diffeomorphism. If $[\omega]=[S]$, the mixed characteristic numbers below agree, but the differential forms need not be identical. Moser's stability theorem gives a global symplectomorphism isotopic to the identity provided that $\omega$ and $S$ are connected by a path of symplectic forms representing their common cohomology class, and equality of the endpoint cohomology classes alone does not guarantee this in general \cite{Moser1965Volume, mcduff2017introduction}.

Physically, $S$ is selected canonically by the topology of the spatial manifold and by the normalization of large gauge transformations. A different $\omega$ is additional data: it may encode a different polarization, a modified pairing of fluxes with generalized currents, or another microscopic response convention. There is no reason for it to equal $S$ unless the physical construction enforces that identification. For example, $\omega=kS$ is locally Darboux-equivalent to $S$, but for $|k|>1$ it has polarization type $(|k|,\ldots,|k|)$ and is not related to $S$ by an integral torus automorphism.

\subsection{Integrality of Hall Conductivities} \label{ssec:GENERALIZED}

With this mathematical machinery in place, we now proceed to establish integrality of a class of generalized Hall conductivities.
See Appendix \ref{app:LEFSCHETZ} for additional details.

Let $\Pi=Q^{-1}$ and define the symplectic contraction operator by
\begin{align}
    \Gamma_S =\frac12\Pi^{MN}\iota_{\partial_M}\iota_{\partial_N},
    \qquad \Gamma_S S = d.
\end{align}
Since $[S]$ and $[\mathcal F/(2\pi)]$ are integral classes,
\begin{align}
    \nu_1=\int_{\mathcal M_{\Theta}}S^{d-1}\wedge\frac{\mathcal F}{2\pi}\in\mathbb Z. \label{sigma 1}
\end{align}
Writing $\Omega=S^d/d!$ for the Liouville volume form and using $S^{d-1}\wedge\mathcal F=(\Gamma_S\mathcal F)S^d/d$ yields:
\begin{align}
    \nu_1=\frac{(d-1)!}{2\pi}
    \int_{\mathcal{M}_{\Theta}}\Omega\, (\Gamma_{S} \mathcal{F}) , \label{sigma 1 expansion}
\end{align}
namely the symplectic trace of the antisymmetric response whose flux-space average is quantized.
When $d=1$, Equation \eqref{sigma 1} reduces exactly to equation \eqref{First Chern number}.

Additionally, for $1\le m\le d$ one has
\begin{align}
    \nu_m=\int_{\mathcal M}S^{d-m}\wedge
    \left(\frac{\mathcal F}{2\pi}\right)^m \in \mathbb Z. \label{top-form topological invariants}
\end{align}
The standard Hall conductivity corresponds to the special case:
\begin{equation}
\sigma_{\mathrm{Hall}} = \frac{\nu_{m = 1}}{2 \pi}.
\end{equation}
More generally, however, the $\nu_{m}$ specify additional integral invariants as obtained by suitable combinations of the current correlator two-point functions. This follows directly from plugging in our expression for $\mathcal{F}$ given in equation (\ref{Berry Curvature}):
\begin{align}
    \frac{1}{2 \pi} \frac{\mathcal{F}_{MN}}{2 \pi} = \sigma_{MN} = i\sum_{r\ne0}
    \frac{\langle\psi_0|\zeta_N|\psi_r\rangle
          \langle\psi_r|\zeta_M|\psi_0\rangle
          -(M\leftrightarrow N)}{(E_r-E_0)^2}.
\end{align}

We emphasize that the quantities $\nu_{m}$ for different $m$ are a priori distinct numbers;
in a particular family they may vanish or satisfy algebraic relations.\footnote{The point is that both $S$ and $\mathcal{F}$ enter this expression. As an illustrative example, consider the surface $\mathbb{CP}^1 \times \mathbb{CP}^1$ with divisors $D_1$ and $D_2$ associated with the two $\mathbb{CP}^1$ factors. The divisors have intersection pairing $D_1 \cdot D_1 = 0$, $D_2 \cdot D_2 = 0$ and $D_1 \cdot D_2 = 1$. Consider the line bundle $\mathcal{L} = \mathcal{O}(a_1 D_1 + a_2 D_2)$, corresponding to $[\mathcal{F} / 2 \pi]$, and $S = \mathcal{K}_{\mathbb{P}_1 \times \mathbb{P}_1} = \mathcal{O}(-2D_1 -2D_2)$ the canonical class of the surface. Observe that simply specifying $c_1(\mathcal{L}) \cdot c_1(\mathcal{K}) = -2 a_1 - 2 a_2$ does not completely specify $\mathcal{L}$; one must also provide $c_1(\mathcal{L})^2 = 2 a_1 a_2$.}
The symplectic linear algebra identity proved in Appendix \ref{app:LEFSCHETZ} gives
\begin{align}
    \nu_m=\frac{(d-m)!}{m!(2\pi)^m}
    \int_{\mathcal{M}_{\Theta}}\Omega\,\Gamma_S^m(\mathcal{F}^m). \label{Lefschetz function}
\end{align}
Because each component $\mathcal F_{MN}$ is a two-point Kubo response, the integrand is an anti-symmetrized product of $m$ linear-response coefficients. It should not be confused with a connected $m$-point current correlator. Nevertheless, we again emphasize that simply specifying a single invariant $\nu_{1}$ does not determine the higher $\nu_{m}$.

\section{Bolza Surface Calculation} \label{sec:BOLZA}

In the previous section we presented a general treatment of topological band theory on a higher-dimensional parameter space. In this section
we consider a concrete example based on $X_2 = \Sigma$ a genus-two Riemann surface. In this case, there are $2g = 4$ independent 1-cycles and the parameter space is topologically a $\widetilde{T}^4 = \mathcal{M}_{\Theta}$. Our goal will be to understand the integrality of the generalized conductivities in this setting. The case of the Bolza surface is of particular interest because it is a genus-two Riemann surface specified as a discrete quotient of the hyperbolic upper half plane. As such, there is a non-abelian generalization of the standard unit cell construction used in the case of a spatial $T^2$. Indeed, a method of images approach was used to construct the relevant current correlation functions for a Kubo-like formula in reference \cite{sun:2024}. Our aim here will be to show that the considerations we presented earlier provide a manifest derivation of this integrality property which is not tied to a particular quotient / method of images construction.

The rest of this section is organized as follows. We begin by reviewing the lattice construction for the Bolza surface, and briefly review
the strategy of reference \cite{sun:2024} to extract the generalized Hall conductivity. Following this, we show how the same quantization result follows more generally from the considerations we presented earlier.

\subsection{Lattice Construction}

\begin{figure}[t!]
    \centering
    \input{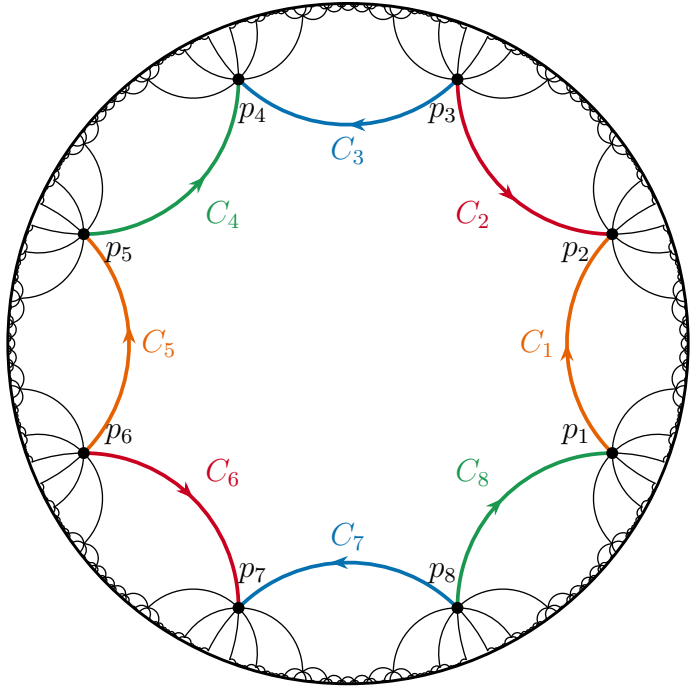}
    \caption{An $\{8,8\}$ tiling of the Poincar\'e disk $\mathbb{D}$. The center octagon is labeled by geodesic sides $C_i$ with respective orientations and lattice points $p_i$. Opposite sides of the same color are identified in order to construct the $g=2$ Bolza surface $\Sigma$.}
    \label{fig:lattice}
\end{figure}

We now review the construction of the Bolza surface by a suitable discrete quotient of a topological disk.\footnote{See \cite{Rayan2021Hyperbolic} for a comprehensive review of both the following construction and many more details on the Bolza surface.} Recall that $T^2$ may be realized from a square unit cell of a lattice where the opposite ends are identified, as shown in \hyperref[fig:torus]{Figure 1}. This process may be extended to all genus $g$ surfaces, but it requires some modifications. First, rather than building the lattice on $\mathbb{R}^2$, we must use the hyperbolic plane $\mathbb{H}$. Specifically, we use the Poincar\'e disk model $\mathbb{D}$, in which $\mathbb{H}$ corresponds to the interior of the complex unit disk. On the Poincar\'e disk, a genus $g$ surface corresponds to a unit cell of a tiling of $\mathbb{D}$ by $4g$-sided polygons with opposite sides identified \cite{Rayan2021Hyperbolic}. For the Bolza surface, we use an $\{8,8\}$ tiling of $\mathbb{D}$, where the first number indicates that each cell is an octagon, and the second that eight cells meet at each lattice point. Finally, the sides of the octagons are chosen to be geodesics of $\mathbb{D}$; see \hyperref[fig:lattice]{Figure 3}.

Geometrically, the Bolza surface (as well as higher genus generalizations) is therefore a quotient of the form:
\begin{equation}
\Sigma = \mathbb{D} / G,
\end{equation}
where $\mathbb{D}$ is a hyperbolic disk, and $G$ is a discrete group. The special case of a torus corresponds to taking $G = \mathbb{Z} \times \mathbb{Z}$, but at higher genus, $G$ is non-abelian.

As is done for $T^2$, the surface is constructed by identifying opposite sides, meaning that we collapse all points $\{p_1,...,p_8\}$ in \hyperref[fig:lattice]{Figure 3} to one point $p_0$. Noting that the first homology group $H_1(\Sigma)$ may be obtained by the abelianization of the fundamental group $\pi_1(\Sigma)$, it is satisfactory to only consider the fundamental group. As shown in \hyperref[fig:lattice]{Figure 3}, we denote the geodesic sides of the center cell by $C_1,...,C_8$, with orientation given by their respective arrows; importantly, under the identification of opposite sides, these all form closed loops through $p_0$, meaning $C_i \in \pi_1(\Sigma, p_0)$. The counter-clockwise loop $C$ about the entire cell is given by
\begin{align}
    C = C_1 C_2^{-1}C_3C_4^{-1}C_5^{-1}C_6C_7^{-1}C_8
\end{align}
and is homotopic to the trivial path, as it may be collapsed to a point, meaning that $C=1$. Once we make the identification of opposite sides, we may write
\begin{align}
    C = C_1 C_2^{-1}C_3C_4^{-1}C_1^{-1}C_2C_3^{-1}C_4 = 1. \label{trivial loop}
\end{align}
With this condition, $\pi_1(\Sigma,p_0)$ is generated by the four loops $C_i$, $1\leq i\leq4$, subject to Equation \eqref{trivial loop}. Since $\Sigma$ is path connected, a different base point gives an isomorphic fundamental group.
\begin{align}
    \pi_1(\Sigma) = \left\langle C_1,C_2,C_3,C_4 \;\middle|\; C_1 C_2^{-1}C_3C_4^{-1}C_1^{-1}C_2C_3^{-1}C_4 = 1 \right\rangle.
\end{align}

Reference \cite{sun:2024} leverages this description in terms of $\Sigma = \mathbb{D} / G$ to study the quantization of the Hall conductivity:
\begin{align}
    \sigma_{ij} = i\sum_{n \neq 0} \frac{\langle \psi_0 | J_j|\psi_n \rangle \langle \psi_n | J_i|\psi_0 \rangle - \langle \psi_0 | J_i|\psi_n \rangle \langle \psi_n | J_j|\psi_0 \rangle}{(E_n - E_0)^2}.
\end{align}
In the case of a genus one Riemann surface, i.e., a $T^2$, it is convenient to work in terms of the Fourier transform of the current. In the higher genus setting, an analogue of this is available based on the representation theory of the non-abelian group $G$. Indeed, one can decompose the external gauge field and current as:
\begin{align}
A &= 2 \pi \theta^{M} \gamma_{M}, \label{eq:ABolza}\\
J &= \zeta^{M} \gamma_{M} \label{eq:JBolza} \\
\end{align}
where the $\gamma_{M}$ correspond to $1$-forms on $\Sigma$. These naturally lift to
$1$-forms $\check{\gamma}^{M}$ defined on $\mathbb{D}$ which
transform in representations of $G$. As found in \cite{sun:2024}, a non-trivial calculation reveals that the Hall conductivity
is still quantized, as follows from the representation theory of $G$.

It is natural to ask whether the same quantization result holds if $\Sigma$ is not represented as a discrete quotient. From
our discussion in the previous sections, we expect this to hold; our aim is to make this explicit by presenting a complementary
treatment which makes manifest the integrality of the Hall conductivity at all steps of the computation.

\subsection{Smooth Construction}

We now show how to recast the calculation of the previous section so that the quantization of the generalized Hall conductivity is
manifest at all stages.

To this end, we observe that our Riemann surface $\Sigma$ is genus two. Now, it is a non-trivial result that genus two Riemann surfaces can be presented as hyperelliptic curves, i.e., as a complex hypersurface of the form:
\begin{equation}
y^2 = f_{5}(x),
\end{equation}
where $(x,y)$ are holomorphic coordinates of $\mathbb{C}^2$, and $f_{5}(x)$ is a degree five polynomial.\footnote{We comment that not all genus $g > 2$ Riemann surfaces can be presented as hyperelliptic curves. This is the famous Schottky problem. Nevertheless, a characterization as a holomorphic curve (all that we really need) is always available.} Indeed, $f_{5}(x)$ has five roots at finite values, and another one ``at infinity'' (upon projectivization of the $x$-plane to a $\mathbb{CP}^1$). Grouping these roots into sets of two, we get branch cuts which can be used to glue together the $y = \sqrt{f_5(x)}$ and $y = - \sqrt{f_5(x)}$ sheets. This naturally yields a genus two Riemann surface. The special case of the Bolza surface corresponds to taking:
\begin{equation}
y^2 = x^5 - x,
\end{equation}
namely, we have a high amount of symmetry in the placement of the roots. Appendix \ref{app:BOLZA} reviews some additional details of the geometry of the Bolza surface, and in particular how to read off the explicit basis of holomorphic 1-forms necessary to apply our previous formulae. The key point is that the basis elements of equations (\ref{eq:JBolza}) and (\ref{eq:ABolza}) can all be presented in terms of holomorphic 1-forms on $\Sigma$.

Everything now goes through as in our general treatment. Appendix \ref{app:BOLZA} constructs an explicit basis of 1-forms
on the Riemann surface, and the Hall conductivity, as well as its generalizations are all manifestly quantized. Since we are dealing with a
genus-two Riemann surface, we have two such invariants, as follows from equation (\ref{Lefschetz function}):
\begin{align}
    \nu_1 &= \frac{1}{2\pi}\int_{\widetilde{T}^4}\Omega\,\Gamma_S(\mathcal{F}) \in \mathbb{Z}\\
    \nu_2 &= \frac{1}{8\pi^{2}}\int_{\widetilde{T}^4}\Omega\,\Gamma_S^2(\mathcal F^2) \in \mathbb{Z},
\end{align}
where we integrate over $\widetilde{T}^4 = \mathcal{M}_{\Theta}$,
the parameter space of the genus-two Riemann surface.

\section{Conclusions \label{sec:CONC}}

We have formulated a flux-space extension of the usual Chern-number description of the Hall response. For a closed oriented spatial manifold $X_{4\ell+2}$, the free middle cohomology carries an alternating unimodular intersection pairing. This pairing induces a canonical principal polarization $S$ on the connected torus of flat middle-degree $U(1)$ backgrounds. A unique gapped ground state over that torus supplies the Berry line bundle, and its antisymmetric adiabatic response is the Berry curvature $\mathcal F$. The resulting integer $\nu_1=\int_{\mathcal{M}_{\Theta}}S^{d-1}\wedge\mathcal F/(2\pi)$ is the flux-space average of the symplectic trace $\Gamma_S\mathcal F$. It reduces to the first Chern number when $\dim\mathcal{M}_{\Theta} = 2$. Higher powers produce additional integral invariants $\nu_m$, which capture further topological information when the parameter torus $\mathcal{M}_{\Theta}$ has dimension greater than two. These quantities are products of linear-response data and need not be mutually independent in a particular model. We have also presented a concrete example based on the Bolza surface which agrees and extends earlier discussions in the literature. In the remainder of this section we discuss some potential avenues for future investigation.

In this work we have focused on the case of a line bundle defined over the moduli space $\mathcal{M}_{\Theta}$. In hyberbolic band theory, it is especially natural to extend this to higher rank bundles. It would be very interesting to use the present formulation to better understand the topological band theory associated with these situations as well.

We have also focused on the torsion free contributions to the middle homology $H_{2 \ell + 1} (X, \mathbb{Z})$. In many cases of interest, there can be additional torsion factors, and these likely contribute additional robust topological observables. Cataloguing these possibilities would likely be quite instructive.

One of the original motivations for this work was to better understand the microscopic degrees of freedom associated with branes which couple to a generalized $p$-form potential. Along these lines, a natural further direction would be to construct interfaces between different bulk Chern-Simons-like theories with distinct levels, with localized degrees of freedom at the interface.

Besides Chern class formulas constructed from the Berry curvature, it is natural to ask whether other observables can be built out of the Berry connection. One natural possibility are Chern-Simons differential forms $\mathrm{CS}_{2m-1}(\mathcal{A})$ (see Appendix \ref{app:CS} for further details). Such forms can be viewed as Chern characters integrated on a space with boundary. It would be interesting to better understand the geometry
of boundary effects in $\mathcal{M}_{\Theta}$, a possibility we leave for future work.

\newpage

\section*{Acknowledgements}

We thank T. Bzdušek, C.L. Kane, J. Maciejko, and C. Sun for helpful discussions. JJH and SR thank the Abdus Salam International
Centre for Theoretical Physics (ICTP) and the organizers of Strings and Geometry
2025 for their hospitality and for a stimulating programme which directly led to the present work.
The work of VC is supported by an NSF Graduate Research Fellowship.
The work of VC and JJH is supported by DOE (HEP)
Award DE-SC0013528 and BSF grant 2022100.
The work of SR is supported by the Natural Sciences and Engineering
Research Council of Canada (NSERC) Discovery Grant program.

\appendix

\section{Symplectic Lefschetz Contraction} \label{app:LEFSCHETZ}

This Appendix proves the linear algebra identity used in Equations \eqref{sigma 1 expansion} and \eqref{Lefschetz function}.
Let $(V,S)$ be a symplectic vector space of dimension $2d$. Define
\begin{align}
    L\alpha=S\wedge\alpha,
    \qquad
    \Gamma_S=\frac12\Pi^{MN}\iota_{\partial_M}\iota_{\partial_N},
    \qquad \Pi=S^{-1},
\end{align}
with the convention $\Gamma_SS=d$. This is the symplectic adjoint of $L$; it is defined by contraction with the Poisson tensor and does not require a Riemannian Hodge star. The standard commutator on an $r$-form is
\begin{align}
    [\Gamma_S,L]\alpha=(d-r)\alpha. \label{Lefschetz commutator}
\end{align}

A form $\alpha_0$ is primitive when $\Gamma_S\alpha_0=0$. The symplectic Lefschetz decomposition expresses every $2m$-form uniquely as
\begin{align}
    \alpha=\alpha_0+L\alpha_1+\cdots+L^m\alpha_m,
\end{align}
where $\alpha_j$ is primitive of degree $2m-2j$. Wedging with $S^{d-m}$ annihilates every term except the scalar term $L^m\alpha_m$; hence
\begin{align}
    L^{d-m}\alpha=\alpha_m L^d. \label{top Lefschetz term}
\end{align}
Repeated use of Equation \eqref{Lefschetz commutator} gives
\begin{align}
    \Gamma_S^m(L^m\alpha_m)
    =\frac{m!d!}{(d-m)!}\alpha_m,
\end{align}
whereas $\Gamma_S^m(L^j\alpha_j)=0$ for $j<m$. Therefore
\begin{align}
    S^{d-m}\wedge\alpha
    =\frac{(d-m)!}{d!\,m!}\,
      \Gamma_S^m(\alpha)\,S^d. \label{symplectic contraction identity}
\end{align}
This pointwise identity applies to differential forms on any symplectic manifold.

Taking $\alpha=\mathcal F^m$ yields
\begin{align}
    S^{d-m}\wedge\mathcal F^m
    =\frac{(d-m)!}{d!\,m!}\,
      \Gamma_S^m(\mathcal F^m)\,S^d.
\end{align}
For $m=1$, it reduces to
\begin{align}
    S^{d-1}\wedge\mathcal F
    =\frac{\Gamma_S\mathcal F}{d}S^d.
\end{align}

\section{Moduli-Space Chern--Simons Characters} \label{app:CS}

The Berry connection also defines secondary invariants on odd-dimensional cycles in $\mathcal M_{\Theta}$. Some care is required: when the Berry line bundle is nontrivial, $\mathcal A$ is not a globally defined one-form, and an integral of $\mathcal A\wedge\mathcal F^{m-1}$ is not by itself a globally defined real number. The invariant object is a differential character, or equivalently its exponentiated holonomy \cite{CheegerSimons1985, Freed1992CS}.

\subsection{Gauge-Invariant Definition}

Let $Z_{2m-1}$ be a closed integral $(2m-1)$-cycle in $\mathcal M_{\Theta}$. In a local trivialization one writes
\begin{align}
    \operatorname{CS}_{2m-1}(\mathcal A;Z_{2m-1})
    =\frac{1}{(2\pi)^m}
      \int_{Z_{2m-1}}\mathcal A\wedge\mathcal F^{m-1}
      \quad\text{in }\mathbb R/\mathbb Z. \label{CS character}
\end{align}
On a nontrivial bundle, the right-hand side is completed by the usual overlap terms between local gauges. The resulting class in $\mathbb R/\mathbb Z$, or its exponential $\exp(2\pi i\operatorname{CS}_{2m-1})$, is gauge invariant.

If $Z_{2m-1}=\partial W_{2m}$ and the Berry bundle with connection extends over $W_{2m}$, then the differential-character Stokes formula is
\begin{align}
    \operatorname{CS}_{2m-1}(\mathcal A;Z_{2m-1})
    =\int_{W_{2m}}\left(\frac{\mathcal F}{2\pi}\right)^m
    \quad \pmod{\mathbb Z}. \label{Chern-Simons Stokes}
\end{align}
Changing the filling changes the right-hand side by the integral of $c_1(\mathcal{L})^m$ over a closed $2m$-cycle, hence by an integer. A cycle need not bound inside $\mathcal M_{\Theta}$ for the differential character to be defined; Equation \eqref{Chern-Simons Stokes} is simply a convenient way to compute it when an extension is available.

\subsection{The \texorpdfstring{$m=1$}{m=1} Case and Pumped Charge}

For $m=1$, Equation \eqref{CS character} is the Berry holonomy
\begin{align}
    \operatorname{CS}_1(\mathcal A;Z_1)
    =\frac{1}{2\pi}\oint_{Z_1}\mathcal A\quad\pmod{\mathbb Z}.
\end{align}
If two loops $Z_{1,i}$ and $Z_{1,f}$ are the oriented boundary of a cylinder $W_2$, then
\begin{align}
    \frac{1}{2\pi}\left(\oint_{Z_{1,f}}\mathcal A
    -\oint_{Z_{1,i}}\mathcal A\right)
    =\int_{W_2}\frac{\mathcal F}{2\pi}\quad\pmod{\mathbb Z}. \label{polarization change}
\end{align}
When one direction of the cylinder is a periodic flux and the other is an adiabatic path, this is the familiar geometric content of the modern theory of polarization \cite{vanderbilt2018berry, Pol_king1993theory, Pol_vanderbilt1993electric, Pol_resta1994macroscopic}.
Only the change along a path, or the holonomy modulo an integer, is gauge invariant.

\subsection{Higher Degree}

For $m>1$, Equation \eqref{Chern-Simons Stokes} transgresses the characteristic class $c_1(\mathcal L)^m$. Its curvature is an anti-symmetrized product of $m$ two-point adiabatic response tensors. This gives a well-defined secondary geometric quantity, but it does not by itself establish a connected $m$-point current observable or a $m$th-order nonlinear response coefficient. Such an interpretation would require a separate microscopic derivation and a specified multi-parameter driving protocol.

The mixed invariant of the main text has a different, though related, role:
\begin{align}
    \nu_m=\left\langle [S]^{d-m}c_1(\mathcal L)^m,[\mathcal{M}_{\Theta}]\right\rangle.
\end{align}
It pairs the same characteristic class with the canonical polarization of the full flux torus. A Chern--Simons character instead records a secondary invariant on an odd cycle. Keeping this distinction prevents the global gauge ambiguity of a local Chern--Simons form from being mistaken for an additional absolute transport coefficient.

\section{Bolza Surface as a Hyperelliptic Curve} \label{app:BOLZA}

In this Appendix we discuss in more detail the geometry of the Bolza surface considered in
section \ref{sec:BOLZA}.

\subsection{Topological Considerations}

We begin with a brief discussion of the topological properties of the Bolza surface. With conventions
as in section \ref{sec:BOLZA}, we may change the basis of generators of $\pi_1(\Sigma)$ such that they correspond to canonical $a$- and $b$-cycles of $H_1(\Sigma)$ following abelianization. Defining new generators
$a_1 = C_3, b_1 = C_4^{-1}, a_2 = C_1 C_2^{-1},$ and $b_2 = C_3 C_4^{-1}C_1^{-1}$, it is easy to check that the trivial loop $C$ is now in canonical form, meaning
\begin{align}
    \pi_1(\Sigma) = \left\langle a_1,b_1,a_2,b_2 \;\middle|\; [a_1,b_1][a_2,b_2]=1 \right\rangle.
\end{align}

After abelianization, these generators give a symplectic basis of $H_1(\Sigma;\mathbb Z)$. The universal coefficient theorem supplies the period-dual basis $\{\alpha_1,\alpha_2,\beta_1,\beta_2\}$ of $H^1(\Sigma;\mathbb Z)$, characterized by:
\begin{equation}\label{canonical basis}
    \int_{a_M} \alpha_N = \delta^{M}\, _{N},  \,\,\, \int_{b_M} \beta_N = \delta_{MN},  \,\,\, \int_{a_M} \beta_N = \int_{b_M} \alpha_N = 0.
\end{equation}

By construction, the cycles have canonical intersections $a_M\cdot a_N=b_M\cdot b_N=0$ and $a_M\cdot b_N=\delta_{MN}$. In the ordered basis $(a_1,a_2,b_1,b_2)$, the intersection matrix is
\begin{align}
    I = \begin{pmatrix} \label{intersection matrix}
0 & 0 & 1 & 0 \\
0 & 0 & 0 & 1 \\
-1 & 0 & 0 & 0 \\
0 & -1 & 0 & 0
\end{pmatrix}.
\end{align}

The Riemann bilinear relation gives, for closed one-forms $\eta$ and $\xi$,
\begin{align}
    \int_{\Sigma}\eta\wedge\xi
    =\sum_{M=1}^2\left(\int_{a_M}\eta\int_{b_M}\xi
    -\int_{b_M}\eta\int_{a_M}\xi\right).
\end{align}
Using equation \eqref{canonical basis}, this yields precisely equation \eqref{symplectic basis} for the ordered cohomology basis $(\alpha_1,\alpha_2,\beta_1,\beta_2)$ \cite{Griffiths1994Principles}. We therefore confirm that the Bolza surface $\Sigma$ admits a basis of $H^1(\Sigma;\mathbb{Z})$ with a canonical symplectic intersection matrix.

\subsection{Hyperelliptic Presentation}

A possible downside of the above construction is that it is not obvious how to give explicit $\alpha$ and $\beta$ forms dual to the $a$ and $b$ cycles. Accordingly, we offer an alternative construction of the Bolza surface that more easily accommodates explicit $\alpha$ and $\beta$ forms. Rather than defining the Bolza surface $\Sigma$ via a tiling of the Poincar\'e disk, we define the Bolza surface as the smooth projective completion of the affine hyperelliptic curve
\begin{align}
    y^2 = x^5 - x \label{hyperelliptic curve}
\end{align}
over $\mathbb{C}$ \cite{Bolza1887, Burnside1893, Cardona2006}. Here, we remind the reader that $x$ and $y$ refer to ambient holomorphic coordinates of $\mathbb{C}^2$, and equation (\ref{hyperelliptic curve}) cuts out a one-complex dimensional subspace.

More explicitly, we compactify the $x$-coordinate, leaving it valued on the Riemann sphere $\mathbb{CP}^1$. Noting that a hyperelliptic curve of genus $g$ is given by $y^2 = P(x)$ for polynomial $P(x)$ with degree $\text{deg}(P) = 2g+1$ \cite{Griffiths1994Principles, Forster1981Lectures}, we confirm that Equation \eqref{hyperelliptic curve} defines a $g=2$ Riemann surface.\footnote{More precisely, for a genus-$g$ hyperelliptic double cover of $\mathbb{CP}^1$ there are $2g+2$ branch points; for $\deg P=2g+1$, one of them lies at infinity.} Reorganizing, we find that the curve is double-valued, $y=\pm \sqrt{x^5-x}$, and hence double-covers the Riemann sphere via the natural projection $\pi : (x,y) \mapsto x $.

\begin{figure}[t!]
  \centering
  \begin{tikzpicture}[line join=round, >=stealth]

    \shade[ball color=gray!40, opacity=0.55] (0,0) circle (3);
    \draw[gray, very thick] (0,0) circle (3);
    \draw[gray, very thick,] (0,0) ++(180:3) arc (180:360:3 and 1);
    \draw[gray, very thick, dashed]         (0,0) ++(0:3)   arc (0:180:3 and 1);
    \draw[gray, very thick, dashed]        (0,3) arc (90:-90:1 and 3);
    \draw[gray, very thick] (0,-3) arc (-90:-270:1 and 3);

    \coordinate (A) at (0,3);
    \coordinate (B) at (-0.95,-0.97);
    \coordinate (C) at (0.95,0.97);
    \coordinate (D) at (-3,0);
    \coordinate (E) at (3,0);
    \coordinate (F) at (0,-3);

    \draw[teal, dash pattern=on 8pt off 6pt]  (A) -- (D);
    \draw[teal, dash pattern=on 8pt off 6pt]  (A) -- (B);
    \draw[teal, dash pattern=on 8pt off 6pt]  (A) -- (C);
    \draw[teal, dash pattern=on 8pt off 6pt]  (A) -- (E);
    \draw[teal, dash pattern=on 8pt off 6pt]  (F) -- (D);
    \draw[teal, dash pattern=on 8pt off 6pt]  (F) -- (B);
    \draw[teal, dash pattern=on 8pt off 6pt]  (F) -- (C);
    \draw[teal, dash pattern=on 8pt off 6pt]  (F) -- (E);
    \draw[teal, dash pattern=on 8pt off 6pt]  (D) -- (C);
    \draw[teal, dash pattern=on 8pt off 6pt]  (C) -- (E);
    \draw[teal, dash pattern=on 8pt off 6pt]  (E) -- (B);
    \draw[teal, dash pattern=on 8pt off 6pt]  (B) -- (D);
    \greatcirclearc[very thick, red, dashed]            {0}{3}{0.95}{0.97}{3}
    \greatcirclearc[very thick, blue]           {0}{-3}{3}{0}{3}
    \greatcirclearc[very thick, green!55!black] {-0.95}{-0.97}{-3}{0}{3}

    \foreach \p/\lbl/\pos in {A/\infty/above, B/-i/below, C/+i/right, D/-1/left, E/+1/above right, F/0/below left}{
      \fill (\p) circle (2pt);
      \node[\pos] at (\p) {$\lbl$};
    }
  \end{tikzpicture}
  \caption{The Riemann Sphere with branch points $\{0,1,-1,i,-i,\infty\}$ inscribing an octahedron shown in light blue. The branch points are paired into three branch cuts, with Branch Cut $\#1$ in dark blue, Branch Cut $\#2$ in green, and Branch Cut $\#3$ in red.}
  \label{fig:sphere}
\end{figure}
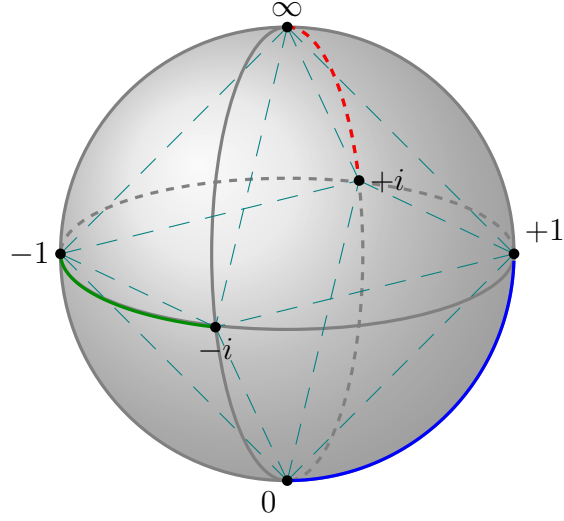

Importantly, the curve possesses six branch points $\{0,1,-1,i,-i,\infty\}$, inscribing an octahedron within the Riemann sphere. We choose to define Branch Cut $\#1$  as the real interval $[0,1]$, Branch Cut $\#2$ as the unit quarter-circle sweeping between $-1$ and $-i$, and finally Branch Cut $\#3$ as the imaginary interval $[i,\infty]$ (See \hyperref[fig:sphere]{Figure 4}).

The $g=2$ topology is realized by considering closed contours traversing both around the branch cuts and through them, thereby connecting the two sheets. One should visualize cross-gluing the two Riemann spheres along the branch cuts. There are six natural cycles of interest:
\begin{enumerate}
  \item Cycle $A_1$ about Branch Cut $\#1$
  \item Cycle $A_2$ about Branch Cut $\#2$
  \item Cycle $A_3$ about Branch Cut $\#3$
  \item Cycle $B_{12}$ connecting Branch Cuts $\#1$ and $\#2$
  \item Cycle $B_{13}$ connecting Branch Cuts $\#1$ and $\#3$
  \item Cycle $B_{23}$ connecting Branch Cuts $\#2$ and $\#3$
\end{enumerate}
Noting that $H_1(\Sigma;\mathbb{Z}) \cong \mathbb{Z}^4$, it is important to identify that only four of the above six cycles are independent. Any of the defined A-cycles may always be expressed as a linear combination of the other two. The same is true for the B-cycles. We will computationally confirm this fact later.

Now that we have identified the relevant cycles, we must construct a basis of harmonic $1$-forms. For a hyperelliptic curve of genus $g$ given by $y^2=P(x)$, $g$ linearly independent holomorphic $1$-forms are
\begin{align}
    \omega_k = \frac{x^{k-1}dx}{y}, \hspace{0.5 cm} \text{for} \hspace{0.3 cm}  k=1,...,g
\end{align}
\cite{Forster1981Lectures}. Thus, $H^{1,0}(\Sigma)$ is spanned by the representatives
\begin{align}
    \omega_1 = \frac{dx}{y} \hspace{0.5 cm} \text{and} \hspace{0.5 cm} \omega_2 = \frac{xdx}{y}.
\end{align}
Noting that the complexified de Rham cohomology group satisfies $H^1(\Sigma)^{\mathbb{C}} \cong H^{1,0}(\Sigma) \oplus H^{0,1}(\Sigma)$ \cite{nakahara2003geometry}, we conclude that $H^1(\Sigma)^{\mathbb{C}}$ is spanned by $\{ \omega_1, \omega_2 ,\bar{\omega}_1, \bar{\omega_2} \}$, and thus the real de Rham cohomology group $H^1(\Sigma)$ is spanned by $\{ \text{Re}[\omega_1], \text{Re}[\omega_2], \text{Im}[\omega_1], \text{Im}[\omega_2] \}$. It follows that for a general genus $g$ surface $\Sigma$, $\text{dim}(H^1(\Sigma)) = 2g$ as noted previously.

We seek to construct a canonical basis of 1-cycles and its canonical dual basis of harmonic $1$-forms. To do so, we first evaluate the integrals of $\omega_i$ over the six cycles. Beginning with the $A$-cycles, and defining the contours to hug the branch cuts tightly, we find:
\begin{equation}
\begin{array}{>{\displaystyle}c @{\qquad} >{\displaystyle}c}
    \int_{A_1} \omega_1 = -2iI & \int_{A_1} \omega_2 = (2i-2i\sqrt{2})I \\[2.0ex]
    \int_{A_2} \omega_1 = (2-\sqrt{2}-i\sqrt{2})I & \int_{A_2} \omega_2 = (-\sqrt{2}+2i-i\sqrt{2})I \\[2.0ex]
    \int_{A_3} \omega_1 = (-2+\sqrt{2}-2i+i\sqrt{2})I & \int_{A_3} \omega_2 = (\sqrt{2} - i\sqrt{2})I
\end{array}
\end{equation}
where
\begin{align}
    I = \int_0^1 \frac{dx}{\sqrt{x-x^5}} = \frac{1}{4}\beta\left(\frac{1}{8},\frac{1}{2}\right) = \frac{\Gamma(\frac{1}{8})\Gamma(\frac{1}{2})}{4\Gamma(\frac{5}{8})}.
\end{align}
It is easily confirmed that integration of either $\omega_1$ or $\omega_2$ about $A_1-A_2-A_3$ gives zero, as predicted, where the relative signs reflect the chosen orientation of the contours around each branch cut. Moving on to the B-cycles, define $B_{12}$ to first traverse along the real axis from $-1$ to $0$ on the positive sheet, $B_{13}$ to first traverse along the imaginary axis from $0$ to $+i$ on the positive sheet, and $B_{23}$ to first traverse along the unit quarter-circle from $-1$ to $+i$ on the positive sheet, and then let all cycles travel in their respective opposite directions on the negative sheet. We find:
\begin{equation}
\begin{array}{>{\displaystyle}c @{\qquad} >{\displaystyle}c}
    \int_{B_{12}} \omega_1 = 2I & \int_{B_{12}} \omega_2 = (-2\sqrt{2}+2)I \\[2.0ex]
    \int_{B_{13}} \omega_1 = (-\sqrt{2}+i\sqrt{2})I & \int_{B_{13}} \omega_2 = (-2+\sqrt{2}-2i+i\sqrt{2})I \\[2.0ex]
    \int_{B_{23}} \omega_1 = (2-\sqrt{2}+i\sqrt{2})I & \int_{B_{23}} \omega_2 = (-\sqrt{2}-2i+i\sqrt{2})I
\end{array}
\end{equation}
As before, we confirm that integration of either $\omega_1$ or $\omega_2$ about $B_{12}+B_{13}-B_{23}$ gives zero, where again the relative signs reflect the chosen orientation of the contours.

One symplectic generating set is $\{a_1,b_1,a_2,b_2\}$, where $a_1=A_1$, $b_1=B_{13}$, $a_2=A_2^{-1}$, and $b_2=B_{23}A_1^{-1}$. It satisfies $a_i\cdot a_j=b_i\cdot b_j=0$ and $a_i\cdot b_j=-b_j\cdot a_i=\delta_{ij}$. To find the period-dual cohomology basis, define $\tilde{\alpha}_1=I^{-1}\operatorname{Im}\omega_1$, $\tilde{\alpha}_2=I^{-1}\operatorname{Im}\omega_2$, $\tilde{\beta}_1=I^{-1}\operatorname{Re}\omega_1$, and $\tilde{\beta}_2=I^{-1}\operatorname{Re}\omega_2$. We then set
\begin{gather}
    \alpha_1 \equiv \frac{1}{8} \Big[   (\sqrt{2}-2)\tilde{\alpha}_1 - (4+3\sqrt{2})\tilde{\alpha}_2 + (2+\sqrt{2})\tilde{\beta}_1 - \sqrt{2} \tilde{\beta}_2   \Big], \label{Canonical Cohomology Basis 1}\\
    \alpha_2 \equiv \frac{1}{8} \Big[   \sqrt{2}\tilde{\alpha}_1 - (2+\sqrt{2})\tilde{\alpha}_2 + \sqrt{2} \tilde{\beta}_1 + (2+\sqrt{2})\tilde{\beta}_2   \Big], \label{Canonical Cohomology Basis 2}\\
    \beta_1 \equiv -\frac{1}{4} \Big[(1+\sqrt{2})\tilde{\beta}_1 + \tilde{\beta}_2 \Big], \label{Canonical Cohomology Basis 3}\\
    \beta_2 \equiv \frac{1}{8} \Big[\sqrt{2}\tilde{\alpha}_1 - (2+\sqrt{2})\tilde{\alpha}_2 + (2+\sqrt{2})\tilde{\beta}_1 - \sqrt{2}\tilde{\beta}_2   \Big]. \label{Canonical Cohomology Basis 4}
\end{gather}
It can be confirmed that this basis satisfies
\begin{equation}
\begin{gathered}
    \int_{a_M} \beta_N = \int_{b_M} \alpha_N = 0, \\
    \int_{a_M} \alpha_N = \int_{b_M} \beta_N = \delta^{M}\, _{N},
\end{gathered}
\end{equation}
and is therefore period-dual to the symplectic homology basis. The displayed periods also verify integrality: a de Rham class lies in the image of $H^1(\Sigma;\mathbb Z)$ precisely when all of its periods are integral.

As detailed previously, the ordered basis $\{\alpha_1,\alpha_2,\beta_1,\beta_2\}$ reflects the symplectic intersection structure of the chosen homology basis and therefore satisfies
\begin{gather}
    \int_{\Sigma} \alpha_M \wedge \alpha_N = \int_{\Sigma} \beta_M \wedge \beta_N = 0, \,\,\, \text{and} \,\,\, \int_{\Sigma} \alpha_M \wedge \beta_N = \delta^{M} \, _{N}.
\end{gather}
Thus, equations (\ref{Canonical Cohomology Basis 1} -- \ref{Canonical Cohomology Basis 4}) give an explicit integral symplectic basis of $H^1(\Sigma;\mathbb Z)$.

\newpage

\bibliographystyle{utphys}
\bibliography{GeneralizedBandTheory}

\end{document}